\documentstyle[aps,twocolumn,epsf,graphics,floats,rotate]{revtex}

\renewcommand{\(}{\left(}
\renewcommand{\)}{\right)}

\begin{document}
\twocolumn[\hsize\textwidth\columnwidth\hsize\csname
@twocolumnfalse\endcsname

\title{Bubble Wall Profiles With More Than One Scalar Field:\\ A Numerical Approach}

\author{%
\hfill 
Peter John$^*$
\hfill\raisebox{21mm}[0mm][0mm]{\makebox[0mm][r]{HD-THEP-98-54}}\\
\hfill\raisebox{21mm}[0mm][0mm]{\makebox[0mm][r]{hep-ph/9810499v2}}%
}
\address{Institut f\"ur Theoretische Physik, Universit\"at Heidelberg, 
Philosophenweg 16, 
D-69120~Heidelberg, Germany
}

\date{\today}

\maketitle

\begin{abstract}
We present a general numerical approach to solve the equations for bubble wall
profiles in models with more than one scalar field and CP violating
phases.  We discuss the algorithm as well as several problems
associated with it and show some profiles for demonstration found with our method.
\end{abstract}

\pacs{11.10.Wx,02.60.Pn,64.60.-i}
\vskip1.5pc]


For the emergence of a baryon asymmetry of the Universe the Sakharov
conditions necessarily demand deviation from thermodynamical
equilibrium. This condition is fulfilled in first order phase
transitions.  They take place via nucleation of bubbles separating the
symmetric from the broken phase. A first order phase transition might
occur at temperatures around the electroweak scale. It turned out that
in the Standard Model (SM) there is no phase transition at all for
Higgs masses larger than 72 GeV
\cite{lattice98}. Baryon number generation at the electroweak scale
therefore requires more complicated models with additional light
scalar fields such as Two-Higgs-doublet models (2HDM), MSSM, NMSSM or
further extensions of the SM. In the MSSM there is a window for
electroweak baryogenesis and an upper bound for the Higgs mass of
about $m_H< 105$ GeV with a light stop of mass $m_{{\tilde
t}_R}<m_{top}$
\cite{Bodeker,CQW,Losada,gitter,Cline}. In the NMSSM the bounds on the Higgs mass are even
weaker \cite{Froggatt,Huber}.
\footnotetext{$^*$e-mail: P.John@thphys.uni-heidelberg.de}
\setcounter{footnote}{1}

Having established the existence of a first order phase transition 
one can start the actual calculation of the baryon asymmetry
itself. There are several mechanisms described in the literature 
\cite{CKN,Rubakov,Davoudiasl,chargino}.  All of them need the
knowledge of the profile of the bubble wall during the phase
transition. The kink ansatz in many situations is a good approximation
but it might be interesting to have a more refined description and to
determine which deviations occur in the presence of potentials
depending on two or more Higgs fields and one or more CP violating
phases \cite{Comelli,Hammerschmitt,Visher,Japaner}.  In fact it turns
out that the most important value in the MSSM is the deviation
$\Delta\beta=max_v[v(\beta(v)-\beta(v_c))]/v_c$ from the straight line
between the minima since the baryon asymmetry is
\cite{Seco,ViljaWagner} proportional to
\begin{equation}
I=\int_{-\infty}^{\infty}dx\frac{v^2(x)}{T_c^2}\frac{d\beta(x)}{dx}
\end{equation}
$\tan\beta=\frac{v_2}{v_1}$, $v^2(x)=v_1^2(x)+v_2^2(x)$. Having the exact profile one can calculate
the baryon asymmetry like \cite{Seco} and investigate the dynamics of expanding bubbles
as in refs. \cite{Laine,Enquist,Moore}.

To determine the bubble wall profile beyond a simple ansatz we have to
solve the equations of motion numerically. In the case of more than
one scalar field this is a highly nontrivial task since simple methods
like overshooting-undershooting which work fine in the case of one
scalar field like overshooting-undershooting fail.  So one has to use
methods which, beginning with an ansatz, converge to the actual
solution. They are sometimes called ``relaxation methods''. Their
practical implementation often is quite nontrivial. The issue of this
letter is to present a working algorithm for the computation of bubble
wall profiles.

We first have to find the equations of motion.  In field theory they
can be derived via Euler Lagrange equations from the Lagrangian
density which has the general form
\begin{equation}
{\cal L} = (D_\mu \Phi_i)^+(D^\mu\Phi_i) + V(\Phi_i,T)
\label{L}\end{equation}
for several Higgs fields $\Phi_i$ (plus phases). Here $D_\mu$ is
a covariant derivative and $V$ denotes the effective
potential. The equations of motion also can be derived
thermodynamically similar to
\cite{Laine}.

Let us consider the MSSM where we may have two dynamical Higgs scalars (or others as
a light right-handed stop \cite{Bodeker,CQW,Losada,gitter,Cline}) plus
one CP violating phase $\theta$.  Using the real neutral field
components $\phi_1$, $\phi_2$ and a relative phase $\theta$ the
corresponding classical (tree level) Higgs potential is
\begin{eqnarray}
V_{tree} & = & \frac{1}{2}m_1^2\phi_1^2 + \frac{1}{2}m_2^2\phi_2^2 +
m_{12}^2\phi_1\phi_2\cos\theta\nonumber\\ & & +
\frac{1}{32}(g^2+{g^\prime}^2)(\phi_1^2-\phi_2^2)^2.
\end{eqnarray}	

In the MSSM the phase $\theta$ also enters  the stop mass matrix.

To describe the phase transition we use the resummed one loop finite
temperature effective potential (see
e.g. ref. \cite{Brignole})
\begin{equation}
V_T = V_{tree} + V_1(T=0) + V_1(T\neq 0) + V_{Daisy}.
\end{equation}

For the bubble wall in the MSSM without CP violation
there has been quite an interesting first numerical approach to solve
the problem of critical bubbles with two Higgs fields \cite{Seco}. In
\cite{Visher} the CP profile has been investigated in the
background of a fixed Higgs profile. In \cite{Japaner} there are
detailed investigations on CP-phases with restriction to a straight
line between the minima.

The Euler Lagrange equations of (\ref{L}) lead to the following
set of coupled second order nonlinear differential equations
\begin{eqnarray}
E_1&:=&\partial_\mu\partial^\mu\phi_1+\phi_1(\partial_\mu\theta)(\partial^\mu\theta)-\frac{\partial
V_T(\phi_1,\phi_2,\theta)}{\partial\phi_1}=0\label{E1}\\
E_2&:=&\partial_\mu\partial^\mu\phi_2-\frac{\partial
V_T(\phi_1,\phi_2,\theta)}{\partial\phi_2}=0\label{E2}\\
E_3&:=&\partial_\mu(\phi_1^2\partial^\mu\theta) -\frac{\partial
V_T(\phi_1,\phi_2,\theta)}{\partial \theta}=0\label{E3}.
\end{eqnarray}
The usual method for the SM case with only one Higgs field is to solve
the corresponding single equation numerically by ``turning around''
the effective potential and dealing $x$ for a time $t$. The problem
then can be regarded as an initial value problem for the negative
potential and one can use an overshooting-undershooting
procedure. This works well since there is only one direction in field
space.

This situation is completely different once there are additional
directions in field space. Again one can consider the analogous
mechanical problem with the turned around potential. Assuming that we
do not have friction, the initial value problem is the same as trying
to shoot a marble from one top of a hill (first minimum) to
exactly the top of the other hill (second minimum) somewhere along
the ridge in such a way that it comes to rest on the top of the second
hill. Small changes in the initial conditions lead to a completely
different shape of the solution. It is, in general, not possible to
know the initial conditions with sufficient accuracy to find the desired
solution.

Hence we have to devise another method.  We here use the method of
minimization of the functional of squared equations of motion.
Constraining eqs. (\ref{E1})-(\ref{E3}) to a stationary wall (domain wall) with
velocity $v_w$ at late time $t$ where the wall is already almost flat
we are left with only one spatial  dimension $x=z-v_wt$
perpendicular to the wall.  Then, solving eqs. (\ref{E1})-(\ref{E3})
means finding field configurations for which
\begin{equation}
S_3=\int_{-\infty}^{\infty}dx \(E_1^2(x)+E_2^2(x)+E_3^2(x)\)\label{Funktional}
\end{equation}
is zero which  is achieved by minimizing $S_3$. This method has also successfully
been used in \cite{Seco} for the critical bubble.  The approach of
\cite{Kusenko} in principle also is a minimization procedure and
therefore the following considerations are also applicable.

Using the minimization method we have to solve a boundary
value problem. Thus we have to use an ansatz for every function for
which we want to find the time development which fulfills the boundary
conditions.

\begin{figure}[t]
\vspace*{-2cm}
\hspace*{-.5cm}
\epsfysize=12cm
\centerline{\epsffile{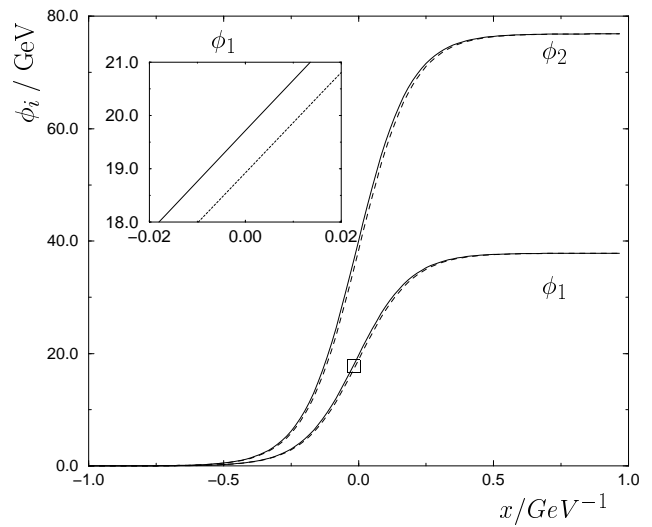}}
\vspace*{-1.5cm} 
\vspace*{-1cm}
\caption{Solution (solid) for the MSSM compared to kink-ansatz (dashed) for $m_A=450$ GeV, $m_Q=350$ GeV, $\tan\beta=2.0$.}
\label{figure1}
\end{figure}

Our procedure works in two steps. The first step is to find an ansatz
which is as close as possible to the exact solution. We will see that
this is a crucial point.  Throughout the literature we find the kink
ansatz for the Higgs fields being quite appropriate. We therefore use
for the $N$ Higgs fields
\begin{equation}
\phi_i^{kink}=\frac{v_i}{2}\(1+\tanh(\frac{x}{L_i})\),
\quad i=1\ldots N\label{kink}
\end{equation}
as our ansatz as well. The $L_i$ are determined by minimizing
(\ref{Funktional}). This first step is cheap in terms of computer time
since it is only a $N$-dimensional minimization and can be performed
very fast. Depending on the actual potential one may also introduce
some further parameters like offsets ${\hat x}_i$ between the fields:
$\phi_i=\frac{v_i}{2}\(1+\tanh(\frac{x}{L_i}+{\hat x}_i)\)$.
Minimizing with respect to a few parameters is a very successfull
first step since it reduces the actual value of (\ref{Funktional})
already sigificantly compared to a general function which only
fullfills the boundary conditions.

In figure \ref{figure1} we show the solution in the MSSM for the
simpler case where we have no CP violation ($\theta(x)=0$) and the
ansatz after this first step of our procedure is just the kink
function $\phi_i(x)=\frac{v_{i}}{2}(1+\tanh(\frac{x}{L_i}))$ with $L_i$ 
now determined. We can recognize that in this case the minimizing
kink function (solid lines) is already quite close to the actual
solution (dashed lines). The shape of the tunneling valley strongly
depends on the CP-odd Higgs mass $m_A$. Small $m_A$ give a larger
mixing with $m_H$ and a sharper bending curve.

Unfortunately the kink ansatz does not serve as a general recipe. For
example CP phases in general cannot be described very well by this
ansatz and one would need further knowledge of their shape in the
wall. For the critical bubble with small radius or potentials with
sharper curves the the ansatz \ref{kink} is not appropriate any
more. One has to choose other types of functions as ansatz and, in
addition, the second step becomes more important.

With this preoptimized ansatz we start the second step of our solving
procedure, the high dimensional minimization. We use
different numerical representations of the action (\ref{Funktional})
which in principle are discretizations of the variables over a grid.
We afterwards compare the results obtained from different
representations. In particular we have to discretize:

i) The space variable $x=z-v_wt$,

ii) the fields $\phi_i(x_j)$,

iii) the derivatives of the fields (first and second derivatives), and

iv) derivatives of explicitely given functions like the effective potential.

ad i) We use a grid of $M$ points ${x_j}, \quad j=1\ldots M$.
According to the special type of solution it is sometimes useful to
transform the integration interval. E.g. for the kink ansatz which
must be integrated over an infinite region it is useful to transform
to $\phi_i(\tau)$ with $\tau=\frac{1}{2}(1+\tanh(\frac{x}{L}))$ where
$\tau\in[0,1]$ which is a finite interval and has the advantage of
many interpolation points where they are needed, in the transition
range of the kink. We have compared this to a equidistant segmentation
and obtained no important difference in the results. By adjusting the
integration interval to the wall thickness $L$ we take care that there
are always enough interpolation points in the wall independent of
$L$. The integration interval ranges between 5 and 10 times $L$.  The
regions outside of this interval do not contribute to
(\ref{Funktional}) any more.

ad ii) The first and quite simple method is just to store the function
values $\phi_i(x_j)$ into an array. To guarantee smoothness here we
need many grid points. We have to store $N\times M$ variables for all
fields.

The second approach is to interpolate the ansatz functions by smooth
functions like splines. There are several types of splines (see
refs. \cite{numrec,Stegun}).  We use cubic splines, which are cubic
polynomials through a given set of points with continous first and
second derivative along the whole interpolation range. Moreover one
can assign the derivatives at the endpoints of the curve which is
useful for setting boundary conditions.  For an interpolation here
only relatively few points $x_j$ are needed. Smoothness is guaranteed
by definition.  The price is an increase in computation time for each
minimization step.  ``Nonlocality'' also has another disadvantage: If a
routine changes one point of the curve for finding a minimum it always
changes a larger region of the curve which may cause eventually larger
changes in (\ref{Funktional}) than useful and maybe, in the worst
case, even a jump into another minimum.  We use both approaches and
compare the results.

ad iii) There are a lot of ways to discretize derivatives as finite
difference equations connecting two or more grid points.  We mostly
use three- or four-point derivatives. Derivatives including more
points may even impair the result since they require more additions
and subtractions which are numerically problematic. Due to quite
different orders of magnitude in the involved numbers they accummulate
numerical errors. For this problem see also
\cite{numrec,Knuth}. For derivative formulae see refs. \cite{numrec,Stegun}

ad iv) Finally the derivatives of explicitely given functions can be
calculated explicitely as well as numerically. This depends on the
problem. For potentials like the MSSM potential we found it to be
sufficent to differenciate numerically.

The minimization itself is accomplished using Powell's method
\cite{numrec} to avoid further derivatives. Minimization parameters
are the values $\phi_i(x_j)$ and $\theta(x_j)$ respectively. So with
$N$ fields defined at $M$ grid points we have a $N\times
M$-dimensional function to be minimized.  Powell's method converges
quadratically and can be used for very high dimensional minimizations
(several hundred dimensions).  Since Powell's method is comparably
slow finding a minimum in an almost flat valley, it might be useful in
such cases to combine it with a Newton's method (converging
quadratically as well) or the like which has a good convergence
behaviour as well. Roughly speaking, Powell's method finds valleys
very fast and Newton's method finds the bottom of the  minimum very
fast. But the converse is not true in general. We also used downhill simplex 
minimization method for comparison which turned out to be much slowlier \cite{numrec}.
Other promising algorithms are ``leap frog'' and related methods \cite{numrec,Moore261}.

Nevertheless there are still several undesired minima which can be
categorized as follows:

a) Some are real solutions to the equations of motion. As an trivial  example
consider $\theta(x)=0,\forall x$ which always is a solution to
(\ref{E1})-(\ref{E3}).

b) Fake minima due to the numerical representation of the
functional. This is a common problem when including derivatives,
discretized as finite differences.

c) Minimization of (\ref{Funktional}) is achieved by solving  $\delta S=0$.
For a squared form $S=a^2$ in addition to the (desired) solution $a=0$
also  pseudo-solutions according to $\delta a=0$ might exist.

\begin{figure}[t]
\vspace*{-2.cm}
\hspace*{-1.5cm}
\epsfysize=12cm
\centerline{\epsffile{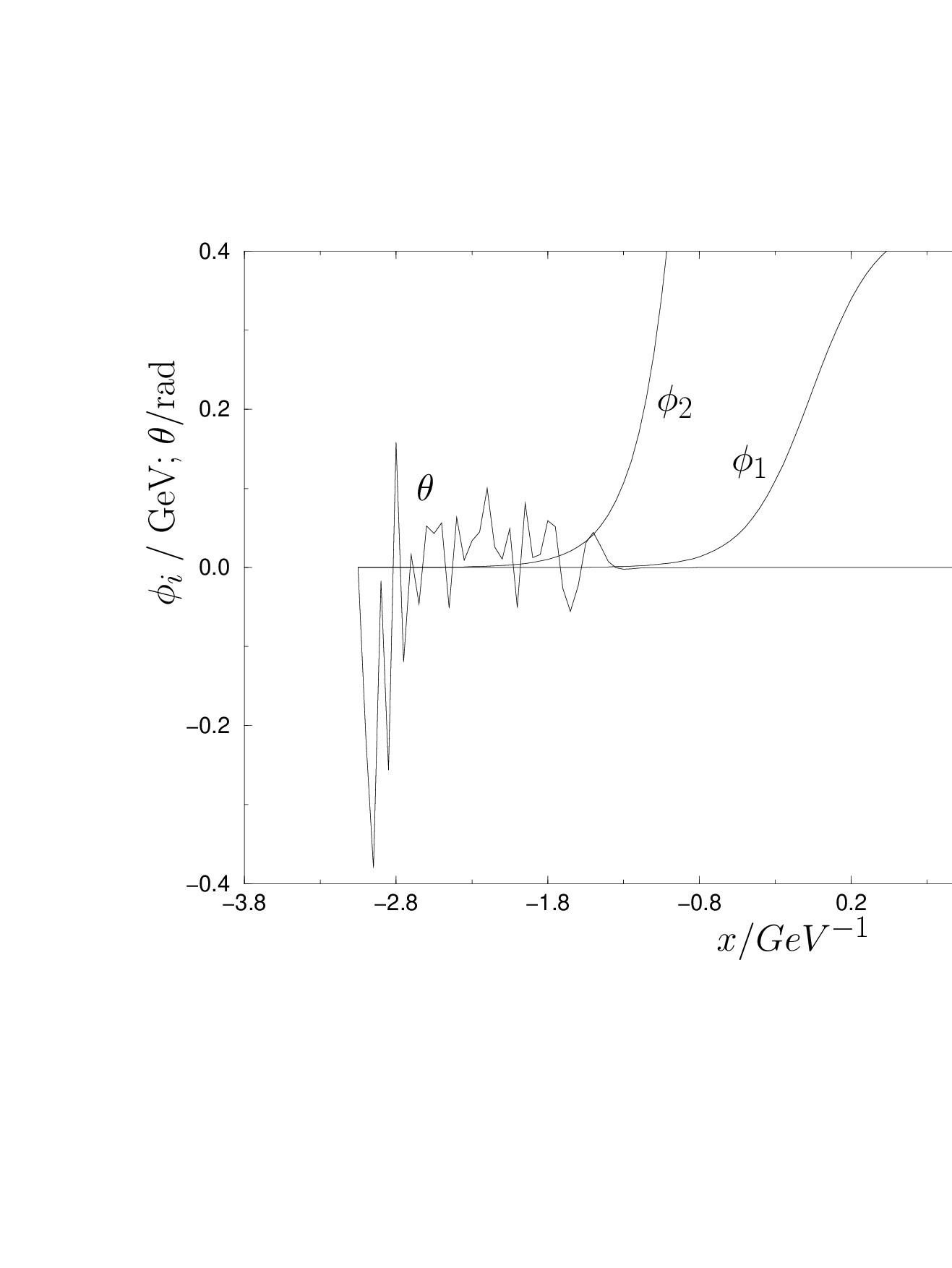}}
\vspace*{-1.5cm} 
\vspace*{-1cm}
\caption{Oscillation of an undesired solution of $\theta(x)$ near the origin,where $\theta$ is small and decouples from $\phi_1$ and $\phi_2$.}
\label{oscillations}
\end{figure}

There can also be combined effects between a), b) and c), e.g. one can
have a pseudo-solution which is only found because of numerical
``fluctuations'' during the minimization procedure. The results often
show oscillations at the outer integration regions (see figure
\ref{oscillations}). We found that this is a combination of a) and b)
as we see from the analytical solution of the third equation of motion
with small $\theta$. Then the third equation decouples and at small
$x$ we have the reduced equation
\begin{equation}
\theta''+\frac{4}{L}\theta'-|m_{12}^2|\tan\beta\theta=0
\end{equation}
which has the solution 
\begin{equation}
\theta(x)=\exp\{-2x/L\}\cos(\omega x/L),
\end{equation}
a damped oscillation with frequency
$\omega=\sqrt{4-L^2|m_{12}^2|\tan\beta}$ and and growing amplitudes
for $x\to -\infty$.

This is what we find numerically. But since our boundary condition
were $\theta'(\pm\infty)$=0 and we took as an ansatz the zero function
$\theta(x)=0$, we find this behaviour to be caused by the minimization
routine.  The routine came through a configuration around
$\theta(x)=0$.  It differs inessentially from the ansatz but together
with the chosen boundary conditions this configuration fits into the
oscillation described above.  Without this pseudo-solution $\theta(x)$
would remain zero for all $x$, as intended.  Fortunately in the region
of interest within the bubble $\theta(x)$ remains zero since the
chosen parameters permit no CP violation there. This is typical for
situations where an angle with vanishing moduli is not well defined
any more. 

Altogether this implies the possible problem that, starting from an
ansatz which is in the vicinity of such an apparent solution, the
algorithm might never converge to the desired exact solution.
Therefore the importance of a good ansatz is obvious. 

What can be done to avoid some of the problems and rate the quality of
a minimum found? First we see that it is important to have a well
prepared ansatz as near as possible to the desired solution to avoid
reaching an unwanted local minimum after the time consuming second
minimization step.  As long as we have energy conservation, as in our
example, one can check the quality of the results easily. It must be
$T-V=0$ or $T/V=1$ and one can check the deviation. Here $V$ is the
QFT effective potential, the mechanical analog has the opposite sign,
$T$ is the kinetic energy density.  It is possible to get results
where $T/V$ deviates only by few percent from $1.0$.  The energy check
is already quite appropriate, it strongly depends on the quality of
the solutions.  It is not useful to add the energy conservation, which
would be fullfilled by the real solution automatically, since it is
another minimum finding task with all problems we have described.
Only in the stationary case $T-V$ could actually be used instead of
(\ref{Funktional}) whereas our method still holds in a more general
case. One can think of further identities that have to be
fulfilled, as in \cite{Kusenko}.  The described special behaviour of
unwanted oscillations can be suppressed by restricting the parameter
space of the minimization procedure at the boundaries. One can also
increase the number of steps in the procedure to improve the precision
step by step for the cost of computing time.

\begin{figure}[t]
\vspace*{-2.0cm}
\hspace*{-.5cm}
\epsfysize=12cm
\centerline{\epsffile{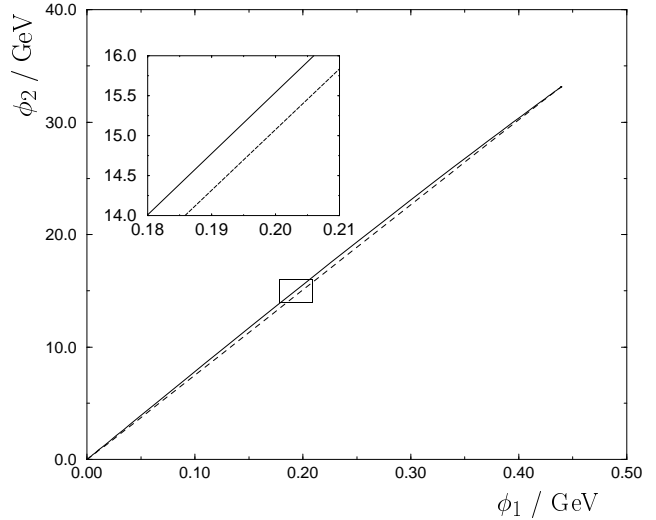}}
\vspace*{-1.5cm} 
\begin{picture}(0,50)(40,0)
\end{picture}
\vspace*{-1cm}
\caption{Unphysically small values of $m_A\approx{\cal O}$(10 GeV) demonstrate the deviation of the solution (solid) from the straight line (dashed).}

\label{figdeltabeta}
\end{figure}

Since the kink solution is a good approximation to the bubble wall
profile equations in a lot of models, we compared our result with
\cite{Seco}. In that paper the radial symmetric equations for the critical bubble were solved.
Thus the solutions have a different shape but in field space there
should be the same behaviour concerning the deviation from the
straight line between the minima. And we indeed can confirm the
results.  $\Delta\beta$ is in the range of $10^{-3}$ to $10^{-2}$ for
$m_A$ around several hundred GeV \cite{Cline,Seco}.  Only for
(experimentally excluded) small values of $m_A\approx {\cal
O}(10\mbox{GeV})$ we find a considerable deviation from a straight
line demonstrating a major deviation from the ansatz due to the
minimization routine (see fig.
\ref{figdeltabeta}). 

\section*{Conclusions and Prospect}
In order to solve the phase transition problem for finding bubble wall
solutions we presented a working algorithm.  Often a simple ansatz
like the kink turns out to be a good approximation. Especially in the
2HDM and the MSSM the deviations from this ansatz are small at least
in some expressions like the surface tension. But for example for the
baryon number asymmetry or the CP violating phase this is not
sufficient since these indeed depend on the small deviation from the
simple ansatz: The latter would give zero for $\Delta\beta$. In some
models this would prevent baryogenesis at all.

With our method we can confirm the results of \cite{Seco} for the one
loop MSSM effective action. The methods presented here can also be
extended to investigate the dynamics of bubble expansion in more
detail. For this we have to go beyond a constant critical temperature
and include reactions of the bubble wall with a background fluid of
particles as in \cite{Moore}. These questions have to be studied
further \cite{future}. An investigation of the NMSSM with three Higgs
fields and a stronger deviation from a straight line in field space is
in progress as well.

\acknowledgments 
I would like to thank Michael G. Schmidt, Mikko Laine, Stephan Huber,
Dietrich B\"odeker, Herbert Nachbagauer and Dirk Jungnickel for useful
discussions and a lot of ideas and suggestions. Thanks also to Marcos
Seco for dicussions during the development of this letter.

Work partially supported by the TMR network {\em Finite Temperature
Phase Transitions in Particle Physics}, EU contract
no. ERBFMRXCT97-0122.


\begin{thebibliography}{12}

\bibitem{lattice98}
K.~Rummukainen,\ M.~Tsypin,\ K.~Kajantie,\ M.~Laine,\ M.~Shaposhnikov,\ hep-lat/9805013

\bibitem{Bodeker} D.~B\"odeker,\  P.~John,\  M.~Laine,\  M.G.\ Schmidt,\  Nucl.\ Phys.\ B497(1997)387

\bibitem{CQW} 
M.~Carena,\  M.~Quir\'os,\  C.E.M.~Wagner,\  Nucl.\ Phys.\ B524(1998)3

\bibitem{Losada} 
M.~Losada,\  hep-ph/9806519

\bibitem{gitter} 
M.~Laine,\  K.~Rummukainen,\  Phys.\ Rev.\ Lett.\ 80(1998)5259,  hep-lat/9804019

\bibitem{Cline} J.M.\ Cline hep-ph/9810267; J.M.\ Cline,  G.D.\ Moore,  hep-ph/9806354 to appear in Phys.\ Rev.\ Lett.; G.D.~Moore hep-ph/9801204, hep-ph/9805264

\bibitem{Froggatt} A.D.~Davies,\ C.D.~Froggatt,\ R.G.~Moorhouse, Phys.\ Lett.\ B372(1996)88;
M.~Pietroni,\ Nucl.\ Phys.\ B402(1993)27

\bibitem{Huber} 
S.J.~Huber,\  M.G.~Schmidt,\  hep-ph/9809506

\bibitem{CKN} 
A.G.~Cohen,\  D.B.~Kaplan,\  A.E.~Nelson,  Ann.\ Rev.\ Nucl.\ Sci.\ 43(1993)27

\bibitem{Rubakov}V.A.~Rubakov and M.E.~Shaposhnikov,
 Usp.\ Fiz.\ Nauk 166(1996)493

\bibitem{Davoudiasl} 
H.~Davoudiasl,\  K.~Rajagopal,\  E.~Westphal,\  Nucl.\ Phys.\ B515(1998)384

\bibitem{chargino} 
J.M.~Cline,\  M.~Joyce,\  K.~Kainulainen,\   Phys.\ Lett.\ B417(1998)79

\bibitem{Comelli} 
D.~Comelli,  M.~Pietroni,  Phys.\  Lett.\  B306(1993)67

\bibitem{Hammerschmitt}
A.~Hammerschmitt,\ J.~Kripfganz,\ M.G.~Schmidt,\ Z.\ f.\ Phys. C64(1994)105

\bibitem{Visher} 
J.~Cline,  K.~Kainulainen,  A.P.~Visher,  Phys.\  Rev.\  D54(1996)2451

\bibitem{Japaner} 
K.~Funakubo,\ hep-ph/9809517;  K.~Funakubo,  A.~Kakuto,  S.~Otsuki,  F.~Toyoda\ hep-ph/9803444, K.~Funakubo, hep-ph/9909517 \&  references therein

\bibitem{Seco} 
J.M.~Moreno,\ M.~Quir\'os,\ M.~Seco,  Nucl.\ Phys.\ B526(1998)489

\bibitem{ViljaWagner} M.~Carena,\ M.~Quir\'os,\ A.~Riotto,\ I.~Vilja,\ C.E.M.~Wagner,\ Nucl.\ Phys.\ B503(1997)387

\bibitem{Laine} 
M.~Laine,\  Phys.\ Rev.\ D49(1994)3847; 
H.~Kurki-Suonio,\   M.~Laine,\  Phys.\ Rev.\ D54(1996)7163,\ Phys. Rev. Lett. 77 (1996) 3951

\bibitem{Enquist} K.~Enqvist,  J.~Ignatius,  K.~Kajantie, K.~Rummukainen,  Phys.\  Rev.\  D45(1992)3415;
J.~Ignatius,  K.~Kajantie,  H.~Kurki-Suonio,  M.~Laine,  Phys.\  Rev.\  D49(1994)3854

\bibitem{Moore} 
G.~Moore,  T.~Prokopec,  Phys.\  Rev.\  D52(1995)7182,\  Phys.\  Rev.\  Lett.\  75(1995)777

\bibitem{Brignole} 
A.~Brignole,\  J.R.~Espinosa,\  M.~Quir\'os,\  F.~Zwirner,\  Phys.\ Lett.\ B(1994)181

\bibitem{Kusenko}
A.~Kusenko, Phys.\ Lett.\ B358(1995)47-50,51-55

\bibitem{numrec} 
W.~Press et al,  {\em Numerical recipes}, Cambridge University Press 1988

\bibitem{Stegun}M.~Abramowitz,\ I.A.~Stegun,\ {\em Handbook Of Mathematical Functions With Formulas, Graphs and Mathematical Tables}, John Wiley \& Sons, 1964,1972

\bibitem{Knuth} 
D.~Knuth,  {\em The Art of Computer Programming}, vol 2, Addison-Wesley, 2nd ed., 1981

\bibitem{Moore261} J.M.~Cline,\ J.R.~Espinosa,\ G.D.~Moore,\ A.~Riotto, hep-ph/9810261

\bibitem{future} 
P.~John,  M.~Laine,  M.G.~Schmidt,  work in progress

\end{thebibliography}
\end{document}